 \documentclass[aps,prl,twocolumn,superscriptaddress,showpacs,floatfix]{revtex4}
\usepackage{graphicx}
\usepackage{dcolumn}
\usepackage{bm}
\begin{document}

\title{Revealing Sub-Surface Vibrational Modes by Atom-Resolved \\
       Damping Force Spectroscopy}

\author{Makoto Ashino}
\affiliation{Institute of Applied Physics and
             Microstructure Research Centre,
             University of Hamburg, 20355 Hamburg, Germany}

\author{Roland Wiesendanger}
\affiliation{Institute of Applied Physics and
             Microstructure Research Centre,
             University of Hamburg, 20355 Hamburg, Germany}

\author{Andrei N. Khlobystov}
\affiliation{School of Chemistry, University of Nottingham,
             University Park, Nottingham NG7 2RD, UK}

\author{Savas Berber}
\affiliation{Physics Department,
             Gebze Institute of Technology,
             41400 Gebze, Kocaeli, Turkey}
\affiliation{Physics and Astronomy Department,
             Michigan State University,
             East Lansing, Michigan 48824-2320, USA}

\author{David Tom\'anek}
\affiliation{Physics and Astronomy Department,
             Michigan State University,
             East Lansing, Michigan 48824-2320, USA}

\date{\today}


\begin{abstract}
We propose to use the damping signal of an oscillating cantilever
in dynamic atomic force microscopy as a noninvasive tool to study
the vibrational structure of the substrate. We present atomically
resolved maps of damping in carbon nanotube peapods, capable of
identifying the location and packing of enclosed Dy@C$_{82}$
molecules as well as local excitations of vibrational modes inside
nanotubes of different diameter. We elucidate the physical origin
of damping in a microscopic model and provide quantitative
interpretation of the observations by calculating the vibrational
spectrum and damping of Dy@C$_{82}$ inside nanotubes with
different diameters using {\em ab initio} total energy and
molecular dynamics calculations.
\end{abstract}




\maketitle




Detecting surface and subsurface vibrations with atomic spatial
resolution is a daunting endeavor. Reported observations of
molecular vibrations by Inelastic Scanning Tunnelling Spectroscopy
(IESTS) require an electrically conducting substrate~\cite{Ho98}.
Atomic Force Microscopy (AFM) experiments involving ultrasonic
oscillations of elastically indented
samples~\cite{{Yamanaka94},{Kolosov98}} can be performed on
electrically insulating systems, but yield subsurface images with
nanoscale resolution at best. Building on the high spatial
resolution and sensitivity of dynamic non-contact
AFM~\cite{{Morita02},{Garcia02}}, we introduce Damping Force
Spectroscopy (DFS) as a non-invasive tool to study subsurface
structure and vibrational modes in complex molecular systems at
the atomic scale. We have chosen carbon nanotube
peapods~\cite{{Smith98},{Jorio08},{Britz06}} consisting of linear
chains of fullerenes enclosed in single-wall carbon nanotubes
(SWNTs)~\cite{Kitaura06} as a prominent example of supramolecular
compounds. Of particular interest are (M@C$_n$)@SWNT peapods
containing M@C$_n$ metallofullerenes, hollow cages of $n$ carbon
atoms surrounding the metal atom M, known for their unusual
electronic transport behavior~\cite{Shimada02}.


Here, we demonstrate that monitoring the damping of an oscillating
AFM tip provides invaluable information not only about topography,
but also the subsurface vibrational modes that have not been
observed before with atomic-scale spatial resolution. Our DFS
studies of (Dy@C$_{82}$)@SWNT indicate that the observed damping
of the tip oscillation depends sensitively on its position and
host tube diameter of (Dy@C$_{82}$)@SWNT, in agreement with
extensive molecular dynamics (MD) studies reported here. Results
of our predictive calculations trace back the observed damping to
the excitation of local vibrational modes by transferring energy
from the oscillating AFM tip. This truly mechanical oscillator
couples to the enclosing nanotube first and subsequently to the
enclosed molecules, revealing their packing structure.



\begin{figure}[t]
\includegraphics[width=0.9\columnwidth]{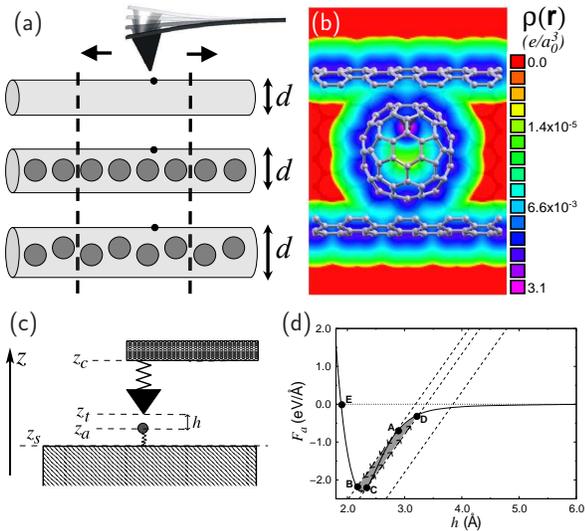}
\caption{(Color)
%
(a) Schematic of an oscillating AFM tip scanning across nanotube
peapods with different diameters $d$ at different filling levels.
%
(b) Total charge density of the optimized Dy@C$_{82}$
metallofullerene encapsulated in the $(18,0)$ nanotube, superposed
with the skeletal structure of the peapod.
%
(c) Schematic of the coupling mechanism between a sharp AFM tip
and a flexible substrate. The cantilever stiffness is represented
by the spring connecting the tip and its suspension. The elastic
substrate is represented by the spring connecting a representative
atom to its surroundings.
(d) Mechanism of energy dissipation during the dynamic AFM
operation in non-contact regime. \label{Fig1}}
\end{figure}

(Dy@C$_{82}$)@SWNT peapods were prepared by encapsulating
Dy@C$_{82}$ metallofullerenes in open-ended SWNTs at a filling
rate of ${\approx}60$\%, as confirmed by high-resolution
transmission electron microscopy and DFS~\cite{EPAPS}. The peapods
were deposited at low coverage onto an insulating flat SiO$_2$
surface of a Si substrate and observed by a home-built dynamic
AFM~\cite{EPAPS}, shown schematically in Fig.~\ref{Fig1}(a). The
AFM, equipped with a commercial Si cantilever (spring constant of
34.3~N/m, eigenfrequency of $159$~kHz) and Si tip (nominal tip
radius of 20~{\AA}), was operated at constant oscillation
amplitude of $21-23$~{\AA} under ultra-high vacuum
($p<1{\times}10^{-8}$~Pa) at low temperature ($T<13$~K).

\begin{figure}[t]
\includegraphics[width=1.0\columnwidth]{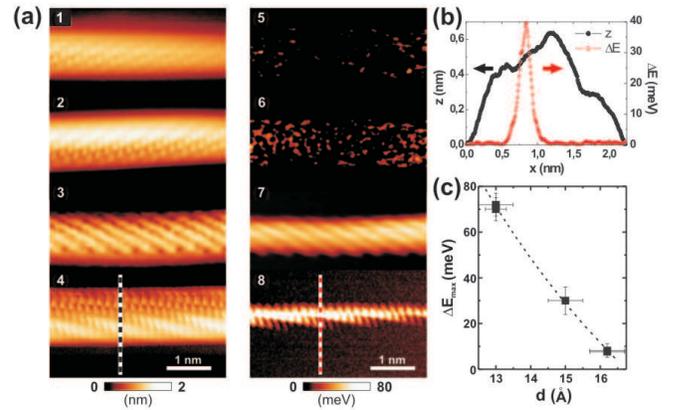}
\caption{(Color) AFM topography and damping in SWNTs and
peapods of different diameter $d$.
(a) Atomically-resolved dynamic AFM observations of the topography
(a:1-4) and damping (a:5-8), defined as energy loss per cycle
${\Delta}E$ and represented by the color-coded
3D contour plots. Results of simultaneous
scans for an empty SWNT with $d=16.2{\pm}0.5$~{\AA} are shown in
(a:1,5), those for peapods with $d=16.2{\pm}0.3$~{\AA} in
(a:2,6), with $d=15.0{\pm}0.5$~{\AA}
in (a:3,7), and with $d=13.0{\pm}0.5$~{\AA}
in (a:4,8).
%
(b) Topographic and damping profiles across the same nanotube
peapod. The topographic data $z(x)$, represented by the black
line, are taken along the black-white dotted trace in (a:4). The
damping data ${\Delta}E(x)$, represented by the red circles, are
taken along the red-white dotted trace in (a:8).
(c) Relationship between nanotube diameter $d$
and the maximum damping energy
${\Delta}E_{max}$. The data points are representative
for tens of tube samples.
The dashed line is a guide to the eye. \label{Fig2}}
\end{figure}

High-resolution AFM topography observations of SWNTs and peapods,
illustrated in Fig.~\ref{Fig1}(a), are shown by
contour plots in Fig.~\ref{Fig2}(a:1-4)~\cite{EPAPS}. Our
topography results indicate that the diameter of the empty SWNT in
Fig.~\ref{Fig2}(a:1) is $d=16.2{\pm}0.5$~{\AA}. Short-range
interatomic interactions, probed by the dynamic AFM, provide
atomic-scale contrast in topography images~\cite{Ashino04}. The
observed atomic arrangement in this tube identifies its chiral
index as $(18,5)$~\cite{Ashino06}. Comparing these results to
those for a peapod of the same diameter, presented in
Fig.~\ref{Fig2}(a:2), we find the atomic-scale topography features
to be amplified with respect to the empty SWNT. This contrast
enhancement
can likely be attributed to local stiffening of the nanotube wall
next to the enclosed Dy@C$_{82}$ molecules.

The optimum packing geometry of fullerenes in peapods is
determined primarily by their diameter ($d_F$) and that of the
enclosing nanotube ($d$)~\cite{Britz06}. We expect an optimum snug
fit
at $d{\approx}d_F+6.4$~{\AA}~\cite{{DT172},{Hodak01}}, with the
center-to-center inter-fullerene distance
$d_{F-F}{\approx}d_F+3.2$~{\AA},
nearly independent of the nanotube diameter. With the diameter
$d_F{\approx}8.0$~{\AA} of Dy@C$_{82}$, a $d{\approx}14.2$~{\AA}
wide SWNT should provide an optimum fit, with the encapsulated
molecules separated by $d_{F-F}{\approx}11.2$~{\AA}. The diameter
$d=13.0{\pm}0.5$~{\AA} of the narrowest peapod observed here,
discussed in Fig.~\ref{Fig2}(a:4), is smaller than optimum. In
this case, the enclosed molecules distort locally
the nanotube wall~\cite{Ashino08},
as seen in Fig.~\ref{Fig2}(a:4). We find the observed undulation
period of $11.5{\pm}0.5$~{\AA} to agree with the above estimated
inter-fullerene distance $d_{F-F}$.


More intriguing than the topography are our results for the
spatial variation of the energy loss ${\Delta}E$ per cycle of the
cantilever oscillation, defined as damping and monitored in
DFS~\cite{EPAPS}. The damping data, shown in
Fig.~\ref{Fig2}(a:5-8), have been recorded simultaneously with the
topography data~\cite{{Morita02},{Garcia02}}. For better
quantitative comparison, we present topography and damping data
for the same system as line plots in Fig.~\ref{Fig2}(b). According
to Fig.~\ref{Fig2}(a:5), almost no damping is observed in an empty
SWNT. In comparison to these observations, Fig.~\ref{Fig2}(a:6)
suggests that damping increases significantly in a peapod of the
same diameter. Decreasing the diameter of the enclosing nanotube
causes an increasingly snug fit of Dy@C$_{82}$ in the peapod. Our
data
in Figs.~\ref{Fig2}(a:6-8) indicate that {\it (i)} ${\Delta}E$
increases, {\it (ii)} the atomic-scale features within ${\Delta}E$
are amplified, and {\it (iii)} helical undulations in ${\Delta}E$
grow in magnitude with decreasing diameter of the enclosing
nanotube. In the case of a snug fit, atomic-scale features in
${\Delta}E$ show a similar or even larger contrast than topography
features. Even though both are spatially correlated, we observe
the strongest damping slightly off the center of the largest
topographic protrusions on-top the encapsulated fullerenes.

The topography images in Figs.~\ref{Fig2}(a) and \ref{Fig2}(b)
show some lateral asymmetry, which may be due to Dy@C$_{82}$
ordering within the peapod or an artifact~\cite{EPAPS}. While we
have taken precautions to avoid artifacts caused by tip-induced
nanotube displacement (we scanned in the axial direction) and
feedback lag (we used a small scan velocity), we can not exclude
an
intrinsically asymmetric shape or doping profile of the AFM tip as
the cause of the asymmetry. Irrespective of this finding, we
conclude that monitoring the damping of the oscillating tip
provides information superior to topography imaging and allows to
discriminate between snug and loose packing of fullerenes inside
the nanotube.


In Fig.~\ref{Fig2}(c) we plot the maximum energy loss per cycle
${\Delta}E_{max}$ observed during a DFS scan as a function of the
host nanotube diameter $d$. The data points in the diagram are
presented for a $(13,6)$ and a $(11,8)$ nanotube with
$d=13.0{\pm}0.5$~{\AA}, a $(19,0)$ nanotube with
$d=15.0{\pm}0.5$~{\AA}, and $(20,1)$ and $(18,5)$ nanotubes with
$d=16.2{\pm}0.5$~{\AA}. Our results indicate a universal
dependence of ${\Delta}E_{max}$ on the diameter $d$ only. From the
narrowest to the widest nanotube, we find the value of
${\Delta}E_{max}$ to decrease from $71{\pm}5$~meV to
$30{\pm}6$~meV and $7.9{\pm}3.0$~meV per cycle of the oscillating
tip.


To uncover the origin of damping in DFS, we performed {\em ab
initio} density functional calculations of the electronic
structure, equilibrium geometry and elastic response of these
systems, along with MD calculations of the dynamical coupling
between the AFM tip and the peapod.

Our total energy calculations indicate that unlike the most stable
C$_2$ isomer of C$_{82}$, the near-spherical isomer of Dy@C$_{82}$
has a different C$_{2v}$ symmetry. This structure is energetically
preferred by at least 0.5~eV over any other isomer and likely
abounds in our samples.
The total charge distribution of Dy@C$_{82}$ inside the $(18,0)$
nanotube, providing optimum enclosure, is shown in
Fig.~\ref{Fig1}(b).


The fundamental concept of an AFM imaging an elastic substrate is
depicted in Fig.~\ref{Fig1}(c). The position $z_t$ of the tip apex
atom depends on the position $z_c$ of the cantilever mount and
the deflection of the elastic cantilever, represented by a
spring. Similarly, an elastic substrate is represented by a
different spring, allowing the position $z_a$ of the closest
surface atom to differ from $z_s$ of the substrate. The force
exerted on the tip by the substrate depends largely on the closest
tip-substrate distance $h=z_t-z_a$.


To elucidate the origin of damping in the dynamical AFM, we first
consider the force $F_a$ exerted on a substrate carbon atom by an
approaching AFM tip at height $h$, shown by the solid line in
Fig.~\ref{Fig1}(d). For a given position $z_c$ of the cantilever
mount, the instantaneous position of the tip $z_t$ and the
substrate atom $z_a$ depends not only on their mutual distance
$h$, which determines the nonlinear interaction force shown by the
solid line, but also on the compensating forces of the strained
cantilever and substrate, shown by the dotted lines, as well as
the history. In the case of a soft cantilever or substrate, as the
AFM tip approaches from far away, an instability occurs at point
A, causing an abrupt decrease in the tip-substrate distance to
point B. During the retraction cycle, a similar instability occurs
at C, causing an abrupt increase in the tip-substrate distance to
D. We note that even the closest-approach point B occurs in the
non-contact regime, since $h_B$ is beyond the equilibrium
tip-substrate separation $h_E$, characterized by $F_a=0$. The
hysteresis corresponding to the shaded area in Fig.~\ref{Fig1}(d),
delimited by A,B,C and D, represents the energy dissipation in the
substrate during an ideal approach-retraction cycle.


To quantitatively analyze, how the energy loss of the oscillating
tip depends on its position and the tube diameter, we performed a
series of MD calculations~\cite{EPAPS}. We induced an initial
perturbation by radially displacing a surface atom by 0.3~{\AA},
corresponding to the distance $h_C-h_E$ in Fig.~\ref{Fig1}(d), and
abruptly releasing it.
The observed energy loss per cycle ${\Delta}E$ depends on the
frequency $h\nu$ of vibrational modes excited by the plucking
process and the likelihood of exciting them. This energy is then
dissipated efficiently owing to the high thermal conductivity of
nanotube systems~\cite{DT130}. Since vibrations excited in the
cantilever do not change with position,
spatial contrast in ${\Delta}E_{max}$ is only due to the
substrate.


In empty SWNTs, we found that plucking excites only the radial
breathing mode (RBM) in the frequency range
$h\nu=17.8-23.1$~meV in the systems considered here. The energy
needed to excite this mode is consistent with ${\Delta}E$ values
seen in Fig.~\ref{Fig2}(a:5). We expect stronger damping in
peapods due to additional modes introduced by the fullerenes.
Ultra-soft vibrations of Dy within the C$_{82}$ cage at
$h\nu=4.1-16.5$~meV are caused by the soft Dy$-$C$_{82}$
interaction and the large mass of Dy. Since these modes are
decoupled from the remaining modes, we will ignore the presence of
Dy in the following discussion. Among the fullerene modes,
librations about the center are the softest, followed by axial and
off-axis motion of the center of mass. Harder fullerene modes
include the RBM at $h\nu=29$~meV, the quadrupolar deformation mode
near $h\nu=62.0$~meV, and higher multipolar modes. Especially in
the case of a snug fit in the peapod, the nanotube and fullerene
modes are strongly mixed.

\begin{figure}[t]
\includegraphics[width=1.0\columnwidth]{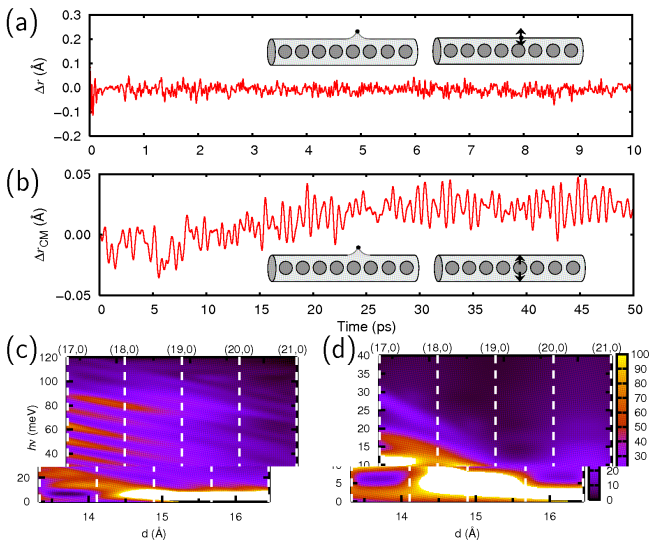}
\caption{(Color) Calculated dynamics and energy dissipation in the
C$_{82}$@SWNT system following an initial 'radial plucking' at the
nanotube surface. Strongest damping occurs when plucking an atom
located axially ${\approx}1$~{\AA} off the center of the nearest
C$_{82}$.
(a) Radial displacement of the plucked atom as a function of
time in C$_{82}$@(18,0).
(b) Radial displacement of the C$_{82}$ center-of-mass as a
function of time in C$_{82}$@(18,0).
(c) Fourier spectrum of the radial displacement of the
initially plucked atom as a function of the nanotube diameter $d$.
(d) Fourier spectrum of the C$_{82}$ center-of-mass as a
function of the nanotube diameter $d$.
The insets in (a) and (b) illustrate the initial
distortion and the quantity monitored. \label{Fig3}}
\end{figure}

For each of the nanotubes studied, we changed the axial position
of the surface atom plucked from on-top a fullerene to in-between
fullerenes, and also studied a position half-way between the two.
Our simulations indicate that damping is strongest for the last
position, due to the possibility of exciting librations
and axial vibrations of the fullerene underneath. This off-center
position, indicated in the inset of Fig.~\ref{Fig3}(a), agrees
with the interpretation of our data in Fig.~\ref{Fig2}(a).

Our MD simulation results for C$_{82}@(18,0)$ are presented in
Figs.~\ref{Fig3}(a) and \ref{Fig3}(b). The radial trajectory  of
the plucked atom, ${\Delta}r(t)$ shown in Fig.~\ref{Fig3}(a), and
that of the closest fullerene molecule, ${\Delta}r_{CM}(t)$ shown
in Fig.~\ref{Fig3}(b), suggest that several modes are excited. To
understand which modes dominate damping, we Fourier analyzed all
trajectories in our study. The Fourier spectra of ${\Delta}r(t)$
are shown in Fig.~\ref{Fig3}(c) and those of ${\Delta}r_{CM}(t)$
in Fig.~\ref{Fig3}(d) for zigzag nanotubes ranging from $(17,0)$
to $(21,0)$. Data for intermediate nanotube diameters have been
obtained by interpolation.

We consider the narrow C$_{82}@(17,0)$ peapod to provide optimum
coupling between the snugly fit fullerene and the enclosing
nanotube. In this case, ${\Delta}E$ and its spatial modulation
should be particularly strong. In contrast to an empty SWNT, the
vibration spectrum of a surface atom in C$_{82}@(17,0)$ is very
rich, as seen in Fig.~\ref{Fig3}(c). Besides low-frequency modes
with $h\nu<20.7$~meV, it contains
modes with equally spaced frequencies spanning the range
$0<h\nu<90.9$~meV. In the wider $(18,0)$ nanotube, the vibration
spectrum is similar, but softer. In the $(19,0)$ and wider
nanotubes, the fullerene-nanotube coupling is reduced drastically,
as fullerenes may rearrange at little energy cost. As seen in
Fig.~\ref{Fig3}(c), decreased fullerene-nanotube coupling causes a
strong intensity drop of the high-frequency modes at
$h\nu>20$~meV. Especially in the narrow nanotubes, our results for
the off-center plucking position in Fig.~\ref{Fig3}(c) differ from
those of other simulations, where the plucked atom is on top of a
fullerene. Since plucking on top a fullerene excites neither
librations nor axial fullerene motion, energy dissipation in this
plucking position is lower. Comparing the results for the two
plucking positions explains the high spatial modulation of
${\Delta}E$ seen in Fig.~\ref{Fig2}(a:8).

The vibration spectrum of the center-of-mass motion of a C$_{82}$
fullerene in the $(17,0)$ nanotube, shown in Fig.~\ref{Fig3}(d),
is rather featureless and much softer than that in
Fig.~\ref{Fig3}(c). The $h\nu{\approx}8$~meV off-axis vibration
mode, which dominates the $(17,0)$ spectrum, red shifts in wider
nanotubes. Owing to the zigzag or helical arrangement of
fullerenes in wider peapods, the off-axis vibrations of the
fullerenes may couple to soft axial vibrations and librations,
which appear in the Fourier spectrum of ${\Delta}r_{CM}$ in
Fig.~\ref{Fig3}(d). These important soft modes, which do not
change ${\Delta}r_{CM}$, modify the spectrum indirectly by
coupling anharmonically to off-axis modes. Especially in the wider
nanotubes, we find the center-of-mass vibrations of C$_{82}$ to be
rather insensitive to the axial position of the plucked atom, thus
contributing very little to the spatial modulation of ${\Delta}E$.

In conclusion, Damping Force Spectroscopy can reveal geometrical
packing and vibrational spectra of subsurface structures at the
atomic scale. We expect that this technique will gain significant
appeal for atomic-level investigations of three-dimensional
supramolecular systems.


We thank Siegmar Roth and Dirk Obergfell for useful discussions
and for sample preparation. We gratefully acknowledge financial
support from the Deutsche Forschungsgemeinschaft and from the
National Science Foundation under NSF-NSEC grant No.~425826 and
NSF-NIRT grant No.~ECS-0506309. Computational resources have been
provided by the Michigan State University High Performance
Computing Center.


\end{document}